\documentclass[structabstract]{aa}
\usepackage{aalongtable}
\usepackage{graphicx,natbib,txfonts,color}
\bibpunct{(}{)}{;}{a}{}{,}

\newcommand{\kms}{km\,s$^{-1}$}

\newcommand{\nodata}{\phantom{0.00}}
\newcommand{\iras}{IRAS\,11472$-$0800}
\begin{document}

\title{ IRAS\,11472$-$0800: an extremely depleted pulsating binary post-AGB star.
\thanks{Based on observations collected at the European Southern
Observatory, Chile. Program ID: 65.L-0615(A), on observations made with the Mercator Telescope, operated on the island of La Palma by the Flemish Community, at the Spanish Observatorio del Roque de los Muchachos and on observations obtained with the HERMES spectrograph, which is supported by the Fund for Scientific Research of Flanders (FWO), Belgium, the Research Council of K.U.Leuven, Belgium, the Fonds National Recherches Scientific (FNRS), Belgium, the Royal Observatory of Belgium, the Observatoire de Gen\`eve, Switzerland and the Th\"uringer Landessternwarte Tautenburg, Germany.
} }

\author{
Hans Van Winckel\inst{1}
\and Bruce J. Hrivnak\inst{2}
\and Nadya Gorlova\inst{1}
\and Clio Gielen\inst{1}
\and Wenxian Lu \inst{2}
}

\offprints{H. Van Winckel, Hans.VanWinckel@ster.kuleuven.be}

\institute{
Instituut voor Sterrenkunde, K.U.Leuven, Celestijnenlaan 200D,
B-3001 Leuven, Belgium 
\and
Department of Physics and Astronomy, Valparaiso University,
Valparaiso, IN 46383, USA
}

\date{Received  / Accepted}

\authorrunning{H. Van Winckel et al.}
\titlerunning{IRAS11472$-$0800}
\abstract{}
{We focus here on one particular and poorly
  studied object, \iras. It is a highly evolved post-Asymptotic Giant
  Branch (post-AGB) star of
  spectral type F, with a large infrared excess produced by thermal
  emission of circumstellar dust. }
{We deploy a multi-wavelength study which includes the analyses of optical and IR
  spectra as well as a variability study based on photometric and spectroscopic time-series.}
{The spectral energy distribution (SED) properties as well as the highly processed
  silicate N-band emission show that the dust in \iras\, is likely
  trapped in a stable disc. The energetics of the SED and the colour
  variability show that our viewing angle is close to edge-on and that the optical
flux is dominated by scattered light.  With photospheric abundances of [Fe/H] = $-$2.7 and
[Sc/H]=$-$4.2, we discovered that \iras\, is one
of the most chemically-depleted objects known to date. Moreover, \iras\, is a
pulsating star with a period of 31.16 days and a peak-to-peak
amplitude of 0.6 mag in $V$. The radial velocity variability is strongly
influenced by the pulsations, but the significant cycle-to-cycle
variability is systematic on a longer time scale, which we interpret as
evidence for binary motion.}
{We conclude that \iras\, is pulsating binary star surrounded by
  a circumbinary disc. The line-of-sight towards the object lies close
the the orbital plane making that the optical light is dominated by scattered
light. \iras\, is one of the most chemically-depleted objects known so far and links
the dusty RV\,Tauri stars to the non-pulsating class of strongly
depleted objects.}

\keywords{Stars: AGB and post-AGB -
 Stars: binaries: general -
 Stars: binaries: spectroscopic -
 Stars: chemically peculiar-
 Stars: evolution}

\maketitle

\section{Introduction}
\label{intro}

Some stars display a peculiar chemical anomaly in their photospheres:
the abundances are depleted and reflect the gas-phase
abundance of the interstellar medium (ISM): chemical species with a
low dust condensation temperature (like Zn and S), are more abundant
relative to elements with a high dust condensation temperature (like
Fe, Ca or the s-process elements).  Among evolved stars, this anomaly
was first recognized in extreme cases like in BD+39~4926
\citep{bond87}, HR\,4049 \citep{lambert88, waelkens91a} and HD\,52961
\citep{waelkens91b,vanwinckel92}, which are objects thought to be in a
post-AGB evolutionary stage. In the latter two objects, the
photospheric Fe abundance is reduced to about 1/60000 times the solar
value while the S and Zn abundances are only slightly less than
solar. HD\,52961 has even a photosphere with more Zn than Fe!  

The process
itself is still poorly understood but the basic ingredients involve a
phase of dusty mass loss. The dust formation process induces the
chemical fractionation as the refractory elements with a high
dust-condensation temperature are preferentially locked up in
solid state dust particles. A gas-dust separation process is followed
by an accretion on the star of the circumstellar gas only, which results in a
photosphere that is coated by a layer of circumstellar gas devoid of
refractory elements \citep{mathis92}.  \citet{waters92} proposed that
the most favourable circumstance for the depletion process to occur is
when (part of the) circumstellar dust is trapped in a stable
circumstellar disc. This allows for a stable environment in which
dust-gas separation and re-accretion can occur. For post-AGB stars, a
stable disc likely implies binarity of the central star, a
proposal that was inspired by the finding that the four extreme cases known
at that time were indeed each part of a binary system \citep{vanwinckel95}

In the last $\sim$15 years, it became clear that this chemical anomaly is very
widespread indeed and not limited to extreme cases. Much milder
depletion patterns are being detected now \citep[e.g.,][and references
therein]{gonzalez97b, giridhar05, maas05, hrivnak08,rao12}. The depleted
objects are not limited to Galactic objects; in the Large
Magellanic Cloud, depleted photospheres are now being found 
as well \citep{reyniers07b, gielen09b}.  In almost all cases the 
spectral energy distributions (SEDs) of the
depleted objects are distinct. The SED is bimodal, with peaks in the
visible and mid-IR, and often includes a prominent near-IR excess
which is interpreted as coming from hot dust in a stable dusty disc
\citep[e.g.,][and references therein]{vanwinckel03}. This characteristic
SED was used to start systematic searches for such systems
\citep{deruyter06}, and in the remainder of this contribution we will
call these objects {\sl disc sources}.

Interferometric studies confirm the very compact nature of the
circumstellar material around the disc sources \citep{deroo06,
  deroo07}, and the infrared spectroscopic data show a very high
processing of the circumstellar dust grains \citep{gielen07,
  gielen08,gielen09b, gielen11}.  Our radial velocity program is still
ongoing, but we indeed confirmed the suspected high binary rate: for
non-pulsating (or low-amplitude pulsating) objects, a binary rate of 100\% was found
\citep{vanwinckel09}.  The companion stars are likely un-evolved
main-sequence stars, which do not contribute significantly to the
energy budget of the objects.  
The orbital periods range from hundred to a few thousand days. The orbits are large enough so that
the actual post-AGB stars fit into their Roche Lobes, but too small to accommodate AGB stars.
The global
picture that emerges is therefore that a binary star evolved in a
system which is too small for a full AGB evolution. During
a badly understood phase of strong interaction, a circumbinary dusty
disc was formed, but the binary system did not suffer dramatic spiral
in. What we now observe is an F-G post-AGB supergiant in a binary
system, which is surrounded by a circumbinary dusty disc
\citep[e.g.][and references therein] {vanwinckel03}. With these observed orbital
characteristics, it is clear that binary interaction processes
dominated their evolution and that these systems represent 
a late phase of binary evolution. The presence of a disc seems to be a
prerequisite for the depletion process to occur, but not all disc
sources are depleted.

A noticeable class of these depleted objects are {\sl dusty} RV\,Tauri
stars \citep{gonzalez97a,gonzalez97b, giridhar98, giridhar00,
  giridhar05, maas05, maas02, vanwinckel98, gielen07}, which occupy
the high luminosity end of the population II Cepheid instability strip
\citep{lloydevans99}. Their SEDs as well as their chemical abundance
patterns suggested that dusty RV\,Tauri stars are also binaries
surrounded by a dusty disc in which the visible star happens to be
located in the population II Cepheid instability strip. Direct
detection of binary motion in these pulsating stars is difficult, but
the binary nature of well-known RV\,Tauri pulsators such as AC\,Her
\citep{vanwinckel98}, EP\,Lyr \citep{gonzalez97a}, RU\,Cen and SX\,Cen
\citep{maas02} is well established. It is important to realise that
there are many RV\,Tauri pulsators without a clear dust excess nor
with a chemical anomaly. There is no observational indication that
also these objects are related to binarity.

In this paper, we focus on the poorly studied object \iras\,
(Table~\ref{tab:init}).  \iras\, was first suspected to be a post-AGB
star based on its large infrared excess as measured by the {\it
  Infrared Astronomical Satellite} (IRAS).  A low-resolution, low
signal-to-noise ratio mid-infrared spectrum was measured with IRAS and
it was classified as showing silicate emission \citep{kwok97}.  It was
unresolved in a mid-infrared imaging survey at a resolution of
$\sim$1.3$\arcsec$ \citep{meixner99}, and it was also spatially
unresolved in a near-infrared (K band) imaging polarimetric survey at
a resolution of 0.4$\arcsec$ \citep{gledhill05}, although it was found
to be highly polarized.  This object was included in a SiO maser
survey but remained undetected \citep{ita01}. The spectrum of the
central star is classified as a F5 Iab.  The strength and shape of its
IR excess, the silicate emission spectrum, as well as the high
polarisation of the unresolved source led to its selection as a
post-AGB disc source candidate \citep{deruyter06}.

\begin{table}
\caption{Specific data on the star \iras\,. \label{tab:init}}
\begin{center}
\begin{tabular}{ll}\hline\hline\rule[0mm]{0mm}{3mm}
 $\alpha$(2000) & 11h\,49m\,48.038s \\
 $\delta$(2000) & -08$^{o}$\,17m\,20.47s \\
 l       & 277.91$^{o}$ \\
 b       & +51.56$^{o}$ \\
 m(v)    & 11.17 \\
 $B-V$     & 0.55 \\ \hline
\end{tabular}
\end{center}
\end{table}


We report here on our detailed study of this particular source.  After
introducing the observations (Sect. 2 ) and the SED (Sect. 3),
 we perform a detailed pulsation analysis (Sect. 4).
The abundance determination is presented in Sect. 5 and the radial
velocity monitoring results are presented in Sect. 6.  We end the
contribution with discussing and highlighting the most important
findings in Sect. 7.


\section{Observations}

\subsection{Visible-band photometry}

Photometric observations of \iras\, were carried out at the
Valparaiso University Observatory (VUO) from 1995 to 2008.  These were
made with the 0.4-m campus telescope and CCD camera using standard
$V$ and $R$ filters.  In the first several seasons the
observations were made primarily with the $V$\, filter and only
occasionally with the $R$\, filter, but beginning in 2000 the $R$\, filter was used regularly.  Unfortunately a problem arose with
the $V$\, filter and no $V$\, data are available from 2000 to
2002.  The object was not observed in 2007.  The $R$\, observations
are on the Cousins photometric system.

Differential photometry was carried out to monitor brightness variations
in \iras\,.  The images were reduced using IRAF \citep{tody93}, with standard bias
and flat field calibration.  An aperture of 11$\arcsec$ was used for
the photometry.  Three comparison stars were monitored, with GSC
05517-00159 used as the main comparison (C$_1$).  All three stars
appear to be constant in brightness,
with C$_1$ and C$_2$ constant at the level of $\pm$0.01 mag based on
their differential measurements.  A total of 75
differential measurements were made with the $V$ filter
($<$$\sigma$$>$=0.012 mag),
 69 with the $R$ filter ($<\sigma>$ =0.010 mag), and 44 ($V-R$) colour indices ($<$$\sigma$$>$=0.014 mag) were
obtained.   
Standardized photometry of the comparison stars and
\iras\, was carried out on two nights, 23 June 1994 at Kitt Peak National
Observatory (KPNO) and 21 May 2009 at the VUO, and the standardized
values are listed in Table~\ref{tab:std_mags}.  The precision in the
two sets of standard magnitudes are $\pm$0.02 and $\pm$0.01 mag, respectively.  
The $R$ and $I$ photometry is on the Cousins system.

One can immediately see that \iras\, varies in brightness from these
two observations (Table~\ref{tab:std_mags}).  This had been recognized earlier and the object has
been assigned the variable star name AF Crt.
It also varies in colour, with the object appearing
bluer when fainter.  This unusual behaviour is confirmed and
discussed below.  The standardized differential magnitudes are listed
in Table~\ref{tab:dif_mags}.

\begin{table*}
\caption{Standard Magnitudes of Program and Comparison
Stars\label{tab:std_mags}} 
\begin{center}
\begin{tabular}{llrrrrr}\hline\hline\rule[0mm]{0mm}{3mm}
Object & GSC ID  & $V$  & $B-V$  & $V-R$
&$R-I$ &Obs Date \\
       &         & mag & mag & mag & mag & \\  \hline\rule[0mm]{0mm}{3mm}
\iras & 05517-00133 &11.60 & 0.82 & 0.51 & 0.56 & 23 Jun 1994 \\
            &    &11.89 & \nodata & 0.34 & \nodata & 21 May 2009 \\
C$_1$  &  05517-00159 &12.92 & 0.61 & 0.31 & 0.33 &  \\
           &     &12.95 &  \nodata  &  0.36 & \nodata & \\
C$_2$ &  05517-00163  &12.94 & 0.66 & 0.37 & 0.38 &  \\
            &    &12.96 &  \nodata  & 0.40 & \nodata &  \\
C$_3$ &  05517-00096  &13.63 & 0.73 & 0.39 & 0.37 &   \\
           &     &13.67 &  \nodata  & 0.44 & \nodata & \\
\hline
\end{tabular}
\end{center}
\end{table*}

\onltab{3}{
\begin{table*}
\caption{\label{tab:dif_mags}
Differential Standard Magnitudes and Colours of IRAS 11472$-$0800 from VUO.}
\begin{tabular}{rrrrrr} \hline\hline\rule[0mm]{0mm}{3mm}
HJD - 2400000 & $\Delta$$V$ & HJD - 2400000 &
$\Delta$$R$ & HJD - 2400000 & $\Delta$($V-R$) \\ 
day           & mag       & day  & mag & day & mag \\
\hline \rule[0mm]{0mm}{3mm}
49840.6193 & $-$1.442 &  49873.5996 & $-$1.589 &  49873.5996&  0.121 \\
49860.6083 & $-$1.676 &  50191.6245 & $-$1.367 &  50191.6212&  0.135\\
49867.6338 & $-$1.627 &  50591.6282 & $-$1.581 &  50591.6245&  0.076\\
49868.6150 & $-$1.609 &  51347.6216 & $-$1.268 &  51347.6216&  0.130\\
49872.5944 & $-$1.502 &  51678.6193 & $-$1.522 &  52732.6257&  0.068\\
49873.5996 & $-$1.468 &  51679.5918 & $-$1.542 &  52741.6658&  0.133\\
50175.6667 & $-$1.576 &  51690.6038 & $-$1.414 &  52763.6147&  0.049\\
50191.6180 & $-$1.233 &  51704.5958 & $-$1.305 &  52780.5934&  0.119\\
50588.6143 & $-$1.511 &  52043.6422 & $-$1.261 &  52781.6047&  0.101\\
50590.5975 & $-$1.502 &  52044.6550 & $-$1.222 &  52782.5879&  0.124\\
50591.6208 & $-$1.505 &  52046.6585 & $-$1.160 &  52786.6118&  0.061\\
50603.6262 & $-$1.129 &  52058.6564 & $-$1.457 &  52787.5943&  0.045\\
50894.6828 & $-$1.357 &  52059.6454 & $-$1.474 &  52789.5986&  0.022\\
50909.7264 & $-$1.042 &  52068.6337 & $-$1.387 &  52792.6031&  0.042\\
50921.6911 & $-$1.605 &  52069.6049 & $-$1.376 &  52795.5974&  0.082\\
50927.6852 & $-$1.440 &  52074.6182 & $-$1.269 &  53098.6218&  0.059\\
50938.6103 & $-$1.220 &  52380.6698 & $-$1.474 &  53110.6166&  0.069\\
50945.6195 & $-$1.114 &  52397.6223 & $-$1.118 &  53119.6202&  0.107\\
50952.6092 & $-$1.609 &  52401.6423 & $-$1.053 &  53143.5859&  0.070\\
51249.7269 & $-$1.371 &  52404.6292 & $-$1.119 &  53152.5923&  0.130\\
51252.7068 & $-$1.238 &  52409.6330 & $-$1.401 &  53159.5890&  0.067\\
51257.7680 & $-$0.990 &  52415.6291 & $-$1.498 &  53160.5946&  0.052\\
51261.6874 & $-$0.999 &  52416.5981 & $-$1.523 &  53170.5992&  0.033\\
51267.6828 & $-$1.538 &  52426.5961 & $-$1.302 &  53463.6248&  0.177\\
51275.6711 & $-$1.445 &  52432.6332 & $-$1.068 &  53469.6741&  0.086\\
51282.7037 & $-$1.264 &  52709.6861 & $-$1.276 &  53474.6341& $-$0.009\\
51299.6350 & $-$1.560 &  52732.6257 & $-$1.322 &  53511.5884&  0.038\\
51308.6129 & $-$1.391 &  52741.6659 & $-$1.209 &  53513.6111&  0.070\\
51318.6663 & $-$0.977 &  52763.6128 & $-$1.382 &  53517.6012&  0.094\\
51325.6084 & $-$1.215 &  52780.5880 & $-$1.032 &  53521.6087&  0.130\\
51326.6159 & $-$1.349 &  52781.6047 & $-$1.018 &  53522.6112&  0.109\\
51337.6498 & $-$1.423 &  52782.5879 & $-$1.014 &  53789.8264&  $-$0.02\\
51347.6216 & $-$1.139 &  52786.6138 & $-$1.213 &  53844.6348&  0.116\\
52732.6258 & $-$1.254 &  52787.5885 & $-$1.255 &  53846.6269&  0.093\\
52741.6656 & $-$1.076 &  52789.5986 & $-$1.266 &  53849.5891&  0.055\\
52763.6165 & $-$1.333 &  52792.6005 & $-$1.291 &  53851.6280&  0.016\\
52780.5988 & $-$0.912 &  52795.5953 & $-$1.298 &  53863.6091&  0.093\\
52781.6046 & $-$0.917 &  52806.6000 & $-$1.158 &  54562.5982&  0.053\\
52782.5879 & $-$0.890 &  53098.6238 & $-$1.100 &  54571.5825&  0.014\\
52786.6098 & $-$1.152 &  53110.6145 & $-$1.413 &  54572.6264&  0.044\\
52787.6002 & $-$1.210 &  53119.6222 & $-$1.244 &  54578.5874&  0.098\\
52789.5986 & $-$1.243 &  53143.5827 & $-$1.379 &  54586.6631&  0.143\\
52792.6057 & $-$1.248 &  53151.6310 & $-$1.215 &  54615.6100&  0.104\\
52795.5995 & $-$1.216 &  53152.5954 & $-$1.169 &  54619.6181&  0.105\\
53098.6197 & $-$1.041 &  53159.5871 & $-$1.135 &            &       \\
53110.6188 & $-$1.344 &  53160.5926 & $-$1.217 &            &       \\
53119.6181 & $-$1.138 &  53170.5972 & $-$1.344 &            &       \\
53143.5891 & $-$1.309 &  53463.6229 & $-$0.942 &            &       \\ 
53152.5892 & $-$1.040 &  53469.6760 & $-$0.894 &            &       \\
53159.5910 & $-$1.068 &  53474.6360 & $-$1.256 &            &       \\
53160.5966 & $-$1.165 &  53511.5905 & $-$1.228 &            &       \\
53170.6013 & $-$1.311 &  53513.5908 & $-$1.222 &            &       \\
53463.6267 & $-$0.765 &  53517.5944 & $-$1.190 &            &       \\
53469.6721 & $-$0.807 &  53521.6067 & $-$1.128 &            &       \\
53474.6321 & $-$1.265 &  53522.6092 & $-$1.093 &            &       \\
53511.5864 & $-$1.189 &  53789.8284 & $-$1.290 &            &       \\
53513.6315 & $-$1.152 &  53844.6368 & $-$1.049 &            &       \\
53517.6079 & $-$1.095 &  53846.6287 & $-$1.054 &            &       \\
53521.6107 & $-$0.997 &  53849.5911 & $-$1.146 &            &       \\
53522.6132 & $-$0.984 &  53851.6301 & $-$1.266 &            &       \\
53789.8244 & $-$1.311 &  53863.6119 & $-$1.301 &            &       \\
53834.6045 & $-$1.180 &  54562.6001 & $-$1.153 &            &       \\
53844.6328 & $-$0.933 &  54571.5843 & $-$1.327 &            &       \\
53846.6250 & $-$0.961 &  54572.6284 & $-$1.301 &            &       \\
53849.5871 & $-$1.092 &  54578.5893 & $-$1.229 &            &       \\
53851.6260 & $-$1.250 &  54580.6033 & $-$1.188 &            &       \\
53863.6063 & $-$1.208 &  54586.6650 & $-$0.970 &            &       \\
54562.5963 & $-$1.099 &  54615.6119 & $-$1.052 &            &       \\
54571.5806 & $-$1.313 &  54619.6200 & $-$0.969 &            &       \\
54572.6245 & $-$1.257 &             &        &            &         \\
54578.5856 & $-$1.131 &             &        &            &         \\
54586.6612 & $-$0.827 &             &        &            &         \\
54612.6158 & $-$1.042 &             &        &            &         \\
54615.6081 & $-$0.948 &             &        &            &         \\
54619.6161 & $-$0.864 &             &        &            &         \\ \hline
\end{tabular}
\end{table*}
}

\subsection{Visible spectroscopy}

The visual spectroscopy includes  high signal-to-noise,
high-resolution optical spectra obtained with UVES mounted on the 8m VLT
telescope within ESO program number 65.L-0615(A). The spectra were
obtained in service mode. We obtained full coverage from 380 nm up to
1000 nm in two spectrograph settings of 900 seconds each.
An image slicer was used to mimic a narrow slit
without compromising the throughput.  We used the dedicated UVES
pipeline to reduce the data in the standard steps for cross-dispersed
echelle spectroscopic data. The final product of our reduction process
is a normalised spectrum of the whole spectral coverage. Measured by 
the standard deviation of continuum windows in the spectrum, 
indicative numbers for the S/N are $\sim$180 at 550 nm, $\sim$100 at 
410 nm and $\sim$160 at 660 nm. 
(An illustration of the quality of the spectra is given in Fig.~\ref{fig:specb}
and Fig.~\ref{fig:spec}.)

Spectral time-series observations of \iras\, were obtained with the Mercator
telescope at the Roque de los Muchachos observatory. We used the
HERMES spectrograph which was specifically designed for this 1.2-m
telescope and which combines a very high throughput with a stable
set-up in a temperature-controlled chamber \citep{raskin11}. The
HERMES spectrograph project developed a dedicated reduction pipeline
which we used for the spectral reduction.  We obtained 49
radial velocity data points over the interval of time from
15 January 2010 to 14 January 2012. The integration times
vary between 1200 and 3600 seconds depending on the local
sky conditions. The object is quite weak for a 1.2 m telescope equipped
with an instrument yielding a spectral resolution of
$\Delta\lambda/\lambda\,\sim \,$85000. The S/N
ratios range from 10 to 30 at 550 nm and at full spectral resolution.

\subsection{Mid-infrared spectroscopy}

We obtained a ground-based N-band low-resolution spectrum with the
TIMMI2 instrument mounted on the 3.6-m telescope of ESO at La Silla,
Chile on 10 March 2004. 
Data reduction was performed in a standard way for the spatially
unresolved source. The chop-nodding observing mode resulted in two
negative and a double positive image of the spectrum on the detector. 
To correct for the variable transmission of the Earth's atmosphere, we
deployed the same method as described by \cite{vanboekel05}, and we obtained a 
spectrum of a calibrator star immediately 
before or after the science target and at very similar airmass. 

\begin{figure}
\resizebox{\hsize}{!}{\includegraphics{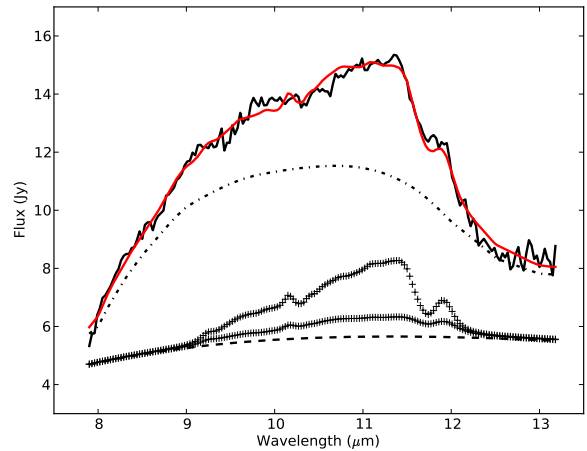}}
\caption{\label{fig:timmi} 
The observed N-band spectrum of \iras is shown as the solid black line. 
The different lines represent our decomposition of
the different dust species contributing to the silicate emission
feature. The dashed line represents the continuum, the dashed-dotted
line the contribution of large grains of amorphous silicates and the +
are large (lower) and small (upper)
grains of forsterite, the Mg-rich end member of the crystalline
olivine. The total model spectrum is depicted in red.}
\end{figure}

The result is a flux-calibrated, low-resolution, N-band spectrum, which is displayed 
in Fig.~\ref{fig:timmi}. The spectrum shows the clear signature of a silicate
emission feature.

\section{Spectral energy distribution}
We used NASA's Astrophysics Data System to complement our own data with fluxes available in the
literature. \iras\ was detected, obviously, by IRAS and additional mid-
and far-infrared data come from AKARI \citep{murakami07}. 
Near-infrared data were obtained from the 2MASS \citep{skrutskie06} and the 
DENIS \citep{epchtein97} projects.  These are listed in Table~\ref{table:sed}.

\begin{figure}
\resizebox{\hsize}{!}{\includegraphics{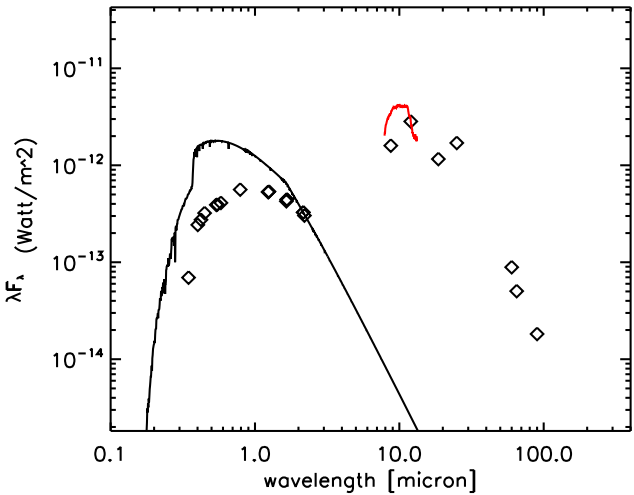}}
\caption{\label{fig:sed} 
SED of \iras\,. 
The observed data points are plotted as open diamonds.
The black line in the optical is the appropriate Kurucz model atmosphere scaled
to the K filter and the full line in the mid-infrared is the N-band TIMMI spectrum. }
\end{figure}

In Fig.~\ref{fig:sed} the SED is displayed. To guide the eye we
matched a Kurucz model atmosphere \citep{castelli04} with appropriate model
parameters ($T_{\mathrm{eff}}$,$\log\,g$ and metallicity) to the K-band flux point. The model parameters
were determined in our spectroscopic analysis (Section 5). Assuming
that the K-band flux is not affected by reddening (interstellar nor
circumstellar) and coming exclusively from the photosphere, the scaled
photospheric model gives us a good measure of the unattenuated
photospheric energy distribution. For the $V$-band, this implies an attenuation
of 1.6 mag. With a galactic latitude of +51.56\degr\, the ISM extinction in
the line of sight towards \iras\, is estimated to be  A$_{V} \sim
0.13$ \citep{drimmel03}.

The SED is unusual as the integrated photospheric model, scaled to the
K band flux points, is about a factor of 6 less luminous
than the integral of the thermal dust emission component.
To investigate the possibility of a strong K reddening, we
estimated the total reddening by minimising the difference between the
scaled photospheric model and the dereddened data.
We assume that the wavelength-dependent reddening follows the ISM
reddening law and we scaled the model fluxes
so that the mean of all the dereddened optical fluxes matches the scaled
photospheric model. A total reddening of E($B-V$)=0.44 +/- 0.01 is
obtained. Also with this reddening, the dereddened fluxes are not high
enough to provide the IR luminosity. We conclude that either the reddening law is
vastly different from the ISM reddening law and includes a strong grey
component or the energy budget of this system is not compatible with a
geometry in which the optical attenuation is redistributed into the infrared. 
\iras\, appears to be an infrared
source with an infrared luminosity which dominates the energetics of
the source!

\begin{table}
\caption{Additional data used to determine the SED as well as the
  energetics of \iras\,.\label{table:sed}} 
\begin{center}
\begin{tabular}{llll}\hline\hline\rule[0mm]{0mm}{3mm}
filter    & $\lambda_c$ & value  & error \\ \hline
m(J)      & 1.25 $\mu$m & 9.657 mag          & 0.023 mag \\
m(H)      & 1.66 $\mu$m& 9.047 mag          & 0.022 mag\\
m(K)      & 2.16 $\mu$m& 8.630 mag          & 0.023 mag\\
IRAS12    & 12 $\mu$m& 11.35 Jy           & 5\% \\
IRAS25    & 25 $\mu$m & 14.18 Jy           & 5\% \\
IRAS60    & 60 $\mu$m & \phantom{0}1.78 Jy & 5\% \\
AKARI-S9W & 8.72 $\mu$m & 4.64 Jy            & 5\% \\
AKARI-L18W& 18.63 $\mu$m& 7.21 Jy            & 5\% \\
AKARI65     & 65.0 $\mu$m  & 1.09 Jy            &10\% \\    
AKARI90     & 90.0 $\mu$m& 0.55 Jy            &10\% \\ \hline
\end{tabular}
\end{center}
\end{table}


\section{Light curves and pulsational variability}

\begin{figure}
\resizebox{\hsize}{!}{\rotatebox{90}{\includegraphics{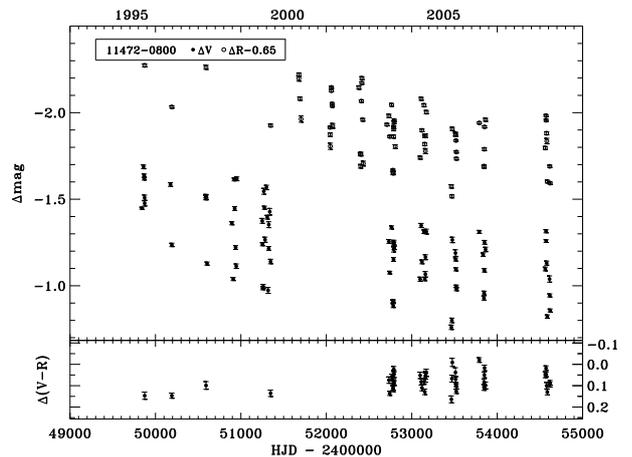}}}
\caption{Differential $V$, $R$, and ($V-R$) light curves of \iras. 
Error bars are included. \label{fig:lcs} }
\end{figure}

The VUO differential light and colour curves are displayed in
Fig.~\ref{fig:lcs}.  They show variations within a season and from
season to season.  There is an approximately monotonic decrease in
median brightness of $\sim$0.5 mag in $V$\, over 13 years, perhaps
reaching a minimum in 2005 and appearing approximately constant or
perhaps increasing by a small amount in brightness from 2005 to 2008.
The $R$\, observations are very few in the early years, but show a
decrease of $\sim$0.35 mag from 2000 to 2005 years.  Both indicate
that the decrease reached a minimum in 2005 and has levelled off or is
showing a slight increase in brightness to 2008.  If this is part of a
long-term periodic variation, it must have a period longer than the
13-year observing interval.  Surprisingly, the average differential
($V-R$) colour of the system from 2003 to 2008 (+0.07 mag) is
bluer by 0.06 mag than it appeared in the few observations made during
the first five years (+0.13 mag), even though the system is fainter on
average by $\sim$0.3 mag between the earlier and later intervals of
time.

\begin{figure}
\resizebox{\hsize}{!}{\includegraphics{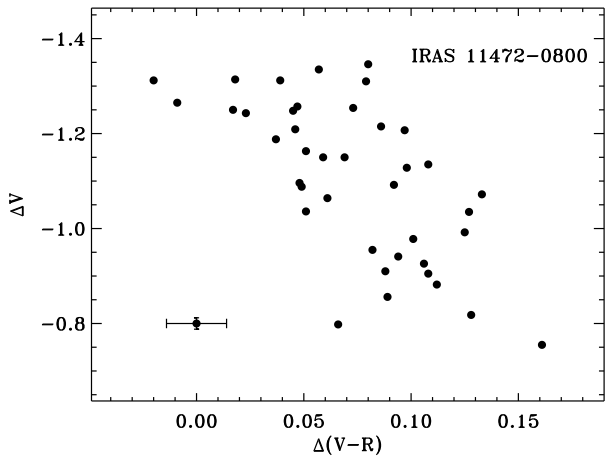}}
\caption{\label{fig:v_color} 
The colour-magnitude diagram for \iras\, from 2003-2008, showing that the object is generally redder when fainter during its pulsational variability.  A standard error bar is shown for reference in the lower left corner.  (Note that we have not included the four points from 1995-1999 when the overall brightness was higher).}
\end{figure}

Brightness variations in an individual season reach a range of up to
0.5$-$0.6 mag and appear to show a cyclical variation.
An examination of the data from 2003 to 2008, when
the overall brightness of the system is about the same, shows that it
is redder when fainter in its cyclical variation, with a range in
($V-R$) colour of $\sim$0.15 mag.  This is shown in
Fig.~\ref{fig:v_color}.  Thus the temperature change is approximately in
phase with the brightness, getting cooler when it is fainter and
hotter when it is brighter.  The $V-R$ colour of the object from
2003 to 2008 varies within a range of 0.12 mag for most of the
observations.

\begin{figure}
\resizebox{\hsize}{!}{\includegraphics{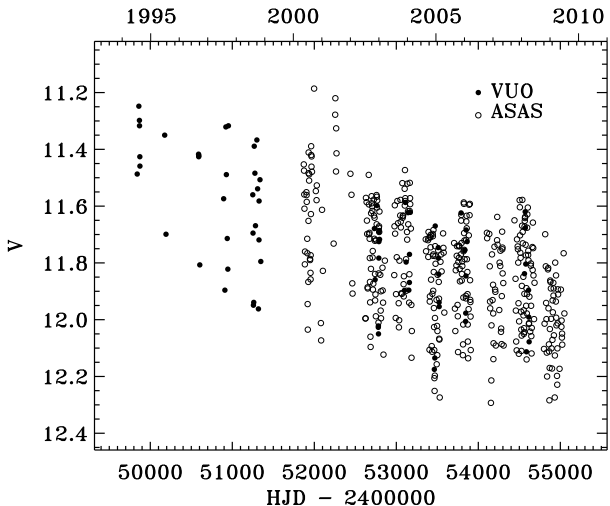}}
\caption{\label{fig:comb_lc} 
Combined $V$ light curve of IRAS 11472$-$0800 based on the VUO and ASAS photometry.}
\end{figure}

Additional $V$\, observations are also available from the All Sky
Automated Survey
\citep[ASAS;][]{pojmanski02}\footnote{http://www.astrouw.edu.pl/asas/}. We used the
measurements made with aperture 2, radius of two pixels (each pixel is
15$\arcsec$ on the sky), and only the good quality data (grade of A or
B).  There were several sets of data for this object, but we
confined ourselves to use the one large set with 420 data points.  The
others were small and were not included since we did not want to
introduce possible offsets by combining the data.  These observations
were made from 2000 through 2009.  A comparison was made with the VUO
data on dates when the observations were made on the same nights or
within one day.  These show an offset, with the ASAS data consistently
brighter by 0.12 mag.  It is not known whether this is the result of a
calibration problem or perhaps the inclusion of an additional star(s)
in the larger ASAS aperture ($r = 30\arcsec$) used for photometry.
The closest bright star is the comparison star that we used, which is
31$\arcsec$ away, and it is possible that some of its light might be
included in the ASAS aperture.  We subtracted this offset from the
ASAS data to combine the two data sets.  This combined $V$ light curve
is shown in Fig.~\ref{fig:comb_lc}.

One can see more clearly in the combined light curve the overall
general decrease in brightness by $\sim$0.6 mag in $V$ from 1995
to 2009, with typical variations within a season of $\sim$0.6 mag
peak-to-peak.  The past seven seasons also appear to show a seasonal
modulation in average system brightness on the order of 0.1 mag.

These light curves were examined for periodicity.  Visual inspection
of the VUO data showed variations that appeared to be consistent with
a cyclical variability but were too few in any season to get a sense
if it is periodic.  However, with the addition of the numerous data
from ASAS, one can see more clearly a cyclical variation of length
30$-$35 d in the $V$ data.

The light curves were formally analyzed for periodicity using the 
Period04 program \citep{lenz05} and in some cases using the CLEAN program \citep{roberts87},
with consistent results.  
We first removed the long-term trend in the light curves by
normalizing each 
of them to their average seasonal values.  Analysis of the VUO data
resulted in well-determined period values of 31.16$\pm$0.01 day ($V$) and 32.18$\pm$0.04 day ($R$).  
Analysis of the ASAS data resulted in a very well-determined period of 31.10$\pm$0.01 day.  
Analyzing the entire $V$ data set (VUO and ASAS), we find a period of 31.14$\pm$0.01 day.
(This changes very slightly to 31.15$\pm$0.01 day if we assign to the
VUO data three times the weight of the ASAS data based on their superior precision.)

\begin{figure}
\resizebox{\hsize}{!}{\includegraphics{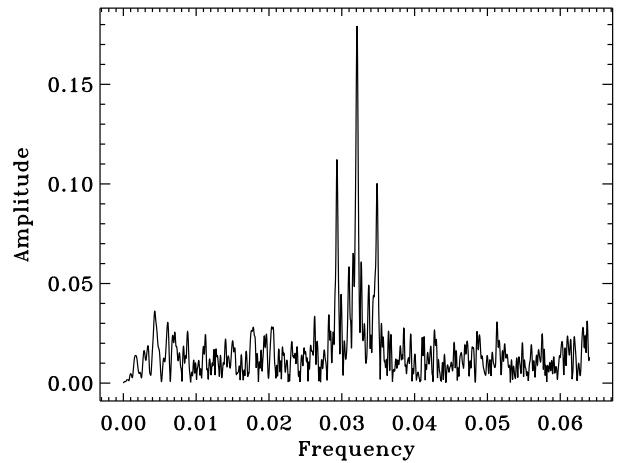}}
\resizebox{\hsize}{!}{\includegraphics{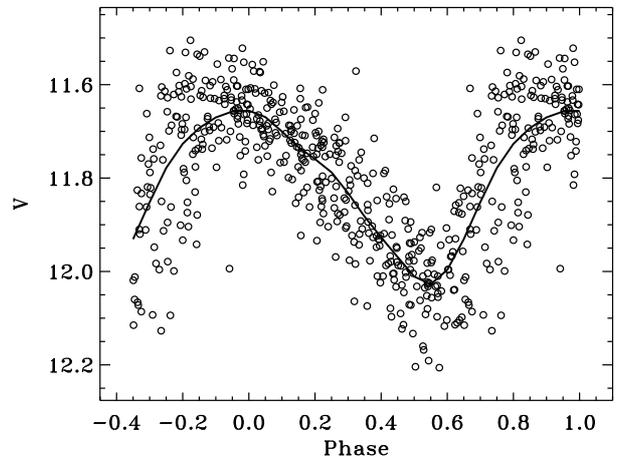}}
\caption{
The frequency spectrum (top) and the phased $V$ light curve (bottom) of
IRAS 11472$-$0800 based on the period of 31.16 d. The spread in the
phased light curve is due in part to the apparent seasonal shifts in
the phase of the light curve (see Fig.~\ref{fig:phase_shift}). 
The solid line is a fit to the data averaged in phase bins. 
(Note that the data from the 2000-2001 and 2001-2002 seasons have been
excluded.) 
\label{fig:freq_phase}}
\end{figure}

An examination of the combined $V$ light curve with the period of
31.14 d showed that the fit was not good for some of the years,
especially the 2000-2001 season.  To investigate this further, we
first analyzed the data season by season for the data sets that had 30
or more points, which were the seasons 2000-2001 to 2008-2009,
excluding 2001-2002 (10 points).  They showed rather similar periods
but changes in the amplitudes and phases.  They also revealed a second
period of about half that of the first, which we interpret as an
attempt to account for the non-sinusoidal shape of the pulsation
curve.  We then, secondly, fixed the period at the value found for the
entire data set and investigated the seasonal values for amplitude and
phase.  We found a good fit for all of the years individually by
allowing for a change in phase.  For all but 2000-2001, the phase
change was within a range of 0.17 $P$.  Examining the VUO
1995-1999 data together with the same period, which we had not
initially included since none of the seasons had 30 or more data
points, we also find a similar phase.  However, the 2000-2001 ASAS
data show a phase change differing from the average of the others by
0.40 $P$\, (12.5 days).  Apparently there occurred a phase or period
change that only affected the 2000-2001 and perhaps the 2001-2002
data, while for the rest of the years the period and phase are
relatively stable.  Determining the period based on the 1995 to 1999
and 2002-2003 to 2008-2009 data combined, but excluding the 2000-2001
and 2001-2002 data, we find $P$ = 31.16$\pm$0.01 day.  The
frequency spectrum based on these data and the data phased to this
period are shown in Fig.~\ref{fig:freq_phase}.  The slight spread in the
light curve shows the effects of the smaller seasonal phase shifts
present in the remaining light curve data.  There are much weaker
secondary periods of 31.55 and 15.57 days, which likely are attempts
to correct for the secondary effects of the remaining phase shifts and
the non-sinusoidal shape of the light curves, respectively.

A recent light curve study of IRAS 11472$-$0800 has been published by
\citet{kiss07} as part of their study of pulsating post-AGB binary
stars.  They determined a similar period of 31.5$\pm$0.6 d based on
the ASAS data from 2000 to 2004 and some unfiltered Northern Sky
Variability Survey
\citep[NSVS;][]{wozniak04}\footnote{http://skydot.lanl.gov/nsvs/nsvs.php}
data from the 1998-1999 and 1999-2000 seasons, and they classified the
object as a Population II Cepheid.  They also cited strong phase
variations in the pulsation phase.  We confirmed the period found by
them based on the smaller data set.  However, the addition of our
earlier data shows more clearly the decrease in system brightness, the
addition of our multicolour observations document the corresponding
change in system colour, and the larger data set results in a higher
precision in the determination of the period.

\begin{figure}
\resizebox{\hsize}{!}{\includegraphics{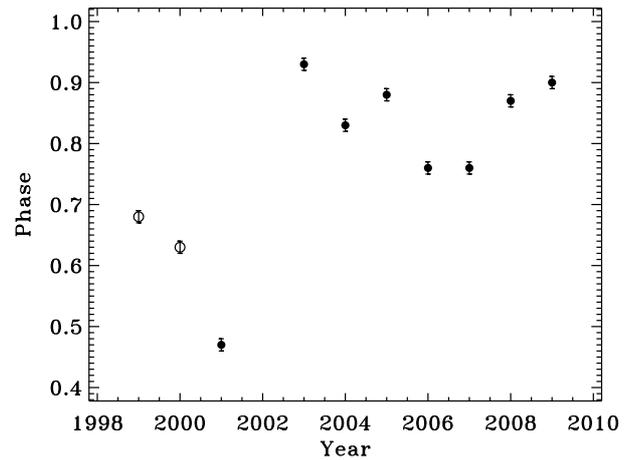}}
\caption{\label{fig:phase_shift} 
 Plot showing the apparent shift in the phase of the seasonal
pulsation curve versus time.  The two open circles are based on the NSVS data.}
\end{figure}

Following the lead of \citet{kiss07}, we show in
Fig.~\ref{fig:phase_shift} the seasonal phase variations in the
pulsational light curves, assuming a fixed period of 31.16 day.  The
phases have an arbitrary zero point; it is the variations in the phase
that are significant.  We folded on $T_o$= 2,449,000JD with a
frequency of 0.032092307. .0We have also included the phase shift
determined by using this period with NSVS data from 1999 to 2000.  The
NSVS data show phase values between the relatively close values found
for the 2002-2003 to 2008-2009 data and the much different 2000-2001
value.

\section{Chemical abundance study}

\begin{figure}
\resizebox{\hsize}{!}{\includegraphics{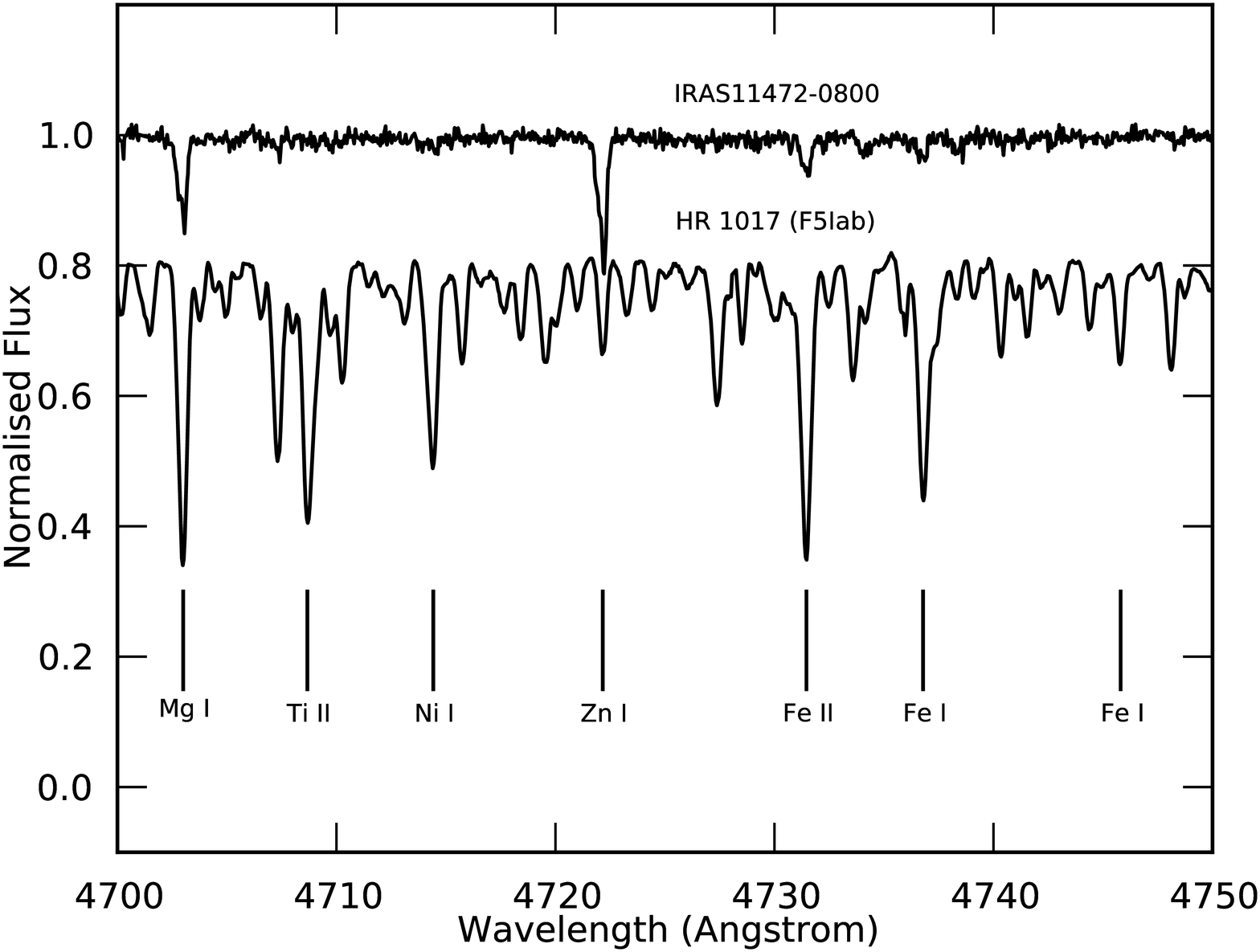}}
\caption{\label{fig:specb} 
Sample of the spectrum of \iras\, showing the effect of depletion on
the spectral lines.  Shown for comparison is HR\,1017, an F5~Iab
supergiant of solar composition.}
\end{figure}

\begin{figure}
\resizebox{\hsize}{!}{\includegraphics{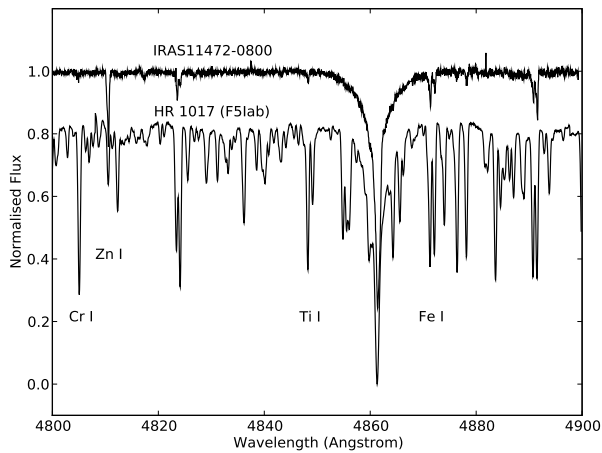}}
\caption{\label{fig:spec} 
Sample of the spectrum of \iras\, in the region of H$\beta$, showing the low abundance of iron 
and the relatively high zinc abundance.  Shown for comparison is HR\,1017, an F5~Iab supergiant.}
\end{figure}

The UVES spectrum with the wide coverage and high S/N
(see examples in Fig.~\ref{fig:specb}, \ref{fig:spec}) was used to
obtain the photospheric chemical abundances. We used the same method
which we have previously described in detail 
\citep[e.g.,][]{vanwinckel00,reyniers03, reyniers07b, hrivnak08}. In short, we use
the measured equivalent widths of small and single atomic lines 
and obtained 
abundances for a wide range of elements by matching
the theoretical equivalent widths to the observed ones. The model
photospheres were obtained from the Kurucz ATLAS9 suite
\citep{castelli04}. We used the 2009 version of 
MOOG \citep{sneden73}\footnote{http://www.as.utexas.edu/$\sim$chris/moog.html}
to determine the abundances.

Model atmosphere parameters were determined in the usual spectroscopic
way. In an iterative process, we fine-tuned the atmospheric model
parameters for which
the abundances are independent of excitation level, ionisation stage,
and relative strength. We strictly limited the analysis to lines with equivalent widths 
smaller than 120 m\AA.  

The low abundances of most elements and relatively high zinc abundance
([Zn/Fe]=+1.8) are
illustrated in the spectra around $\lambda 4720 \AA$ and around H$\beta$,
which are depicted in Figs.~\ref{fig:specb} and
\ref{fig:spec}, respectively. The comparison star is HR1017 which
has similar spectral type, but solar abundances. The final abundances
of \iras\, are
given in Table~\ref{tab:tab_abund}. In the different columns we list
the ion, the number of used lines, the mean equivalent width, the
obtained abundance, and the line-to-line scatter. In the other columns
the relative value with respect to the Sun is given as well as the
dust condensation temperatures from \cite{lodders03}. The latter
are computed in equilibrium with a solar mixture and
under a constant pressure of 10$^{-4}$ atm.

\begin{table}
\caption{Abundance results for \iras. The following model atmosphere
was adopted: $T_{\mathrm{eff}}$\,=\,5750\,K, $\log g$\,=\,1.0,
$V_t$\,=\,4.5\,km\,s$^{-1}$,
and [M/H]\,=\,$-$2.5. \label{tab:tab_abund}}
\begin{tabular}{lrrrrrrr} \hline\hline
Ion & N &
W$_{\lambda}$&$\log\epsilon$&$\sigma_{\rm ltl}$&[X/H]&$\log\epsilon_{\odot}$\tablefootmark{a}&T$_{\rm cond}$\tablefootmark{b} \\ \hline
C\,{\sc i  }  &  13 & 47  &  7.99 &  0.12 & $-$0.44 & 8.43 &   40\\
O\,{\sc i  }  &   3 & 40  &  8.59 &  0.10 & $-$0.10 & 8.69 & 180 \\
Na\,{\sc i  } &   5 & 14  &  5.28 &  0.12 & $-$0.96 & 6.24 &  958\\
Mg\,{\sc i  } &   2 & 67  &  5.45 &  0.03 & $-$2.15 & 7.60 & 1336\\
Al\,{\sc i  } &   1 & 139 &  3.50 & ...   & $-$2.95 & 6.45 & 1653\\
Si\,{\sc i  } &   2 & 10  &  6.34 &  0.24 & $-$1.17 & 7.51 & 1310\\
S \,{\sc i  } &   6 & 28  &  6.67 &  0.11 & $-$0.45 & 7.12 &  664\\
Ca\,{\sc i  } &   4 & 16  &  3.62 &  0.10 & $-$2.72 & 6.34 & 1517\\
Sc\,{\sc ii } &   1 & 34  &  $-$1.09 & ...& $-$4.24 & 3.15 & 1659\\
V\,{\sc ii }  &   1 & 34  &  1.64 &  0.16 & $-$2.29 & 3.93 & 1582\\
Cr\,{\sc i  } &   2 & 61  &  2.68 &  0.01 & $-$2.96 & 5.64 & 1296\\
Mn\,{\sc i  } &   4 & 22  &  3.42 &  0.11 & $-$2.01 & 5.43 & 1158\\
Fe\,{\sc i  } &  43 & 28  &  4.79 &  0.15 & $-$2.71 & 7.50 & 1334\\
Fe\,{\sc ii } &  18 & 40  &  4.87 &  0.14 & $-$2.63 & 7.50 & 1334\\
Cu\,{\sc i  } &   1 & 11  &  2.60 &  ... & $-$2.39 & 4.19 & 1352\\
Zn\,{\sc i  } &   3 & 80  &  3.67 &  0.13 & $-$0.89 & 4.56 &  726\\
Ba\,{\sc ii } &   3 & 32  &$-$1.15&  0.08 & $-$3.33 & 2.18 & 1455\\
\hline
\end{tabular}
\tablefoot{
\tablefoottext{a}{\citet{asplund09}}
\tablefoottext{b}{\citet{lodders03}}}
\end{table}

\section{Radial Velocity Analysis}

The radial velocities are based on the spectral time series obtained
with the HERMES spectrograph \citep{raskin11}. The extreme depletion
means that only very few lines are strong enough to be detected in these
spectra of low signal-to-noise (S/N varies between 10 and 30 at
wavelength of maximal spectral throughput).  
Cross-correlating the spectrum with a spectral mask,
tailored on the basis of the extensive list of weak lines present in the chemical peculiar
star and measured on the UVES high S/N spectrum, failed; 
too few strong lines are present in the
spectrum and only those lines are detectable in the low S/N spectra.

We therefore based our radial velocity determination only on the
strong Mg\,{\rm I} lines ($\lambda$ = 5167.32, 5167.487, 5172.684, 5169.296 and 5183.604 \AA),
and adopted the quality criterion 
so that the S/N of the spectrum, the cross-correlation width, standard deviation and depth should be within 3 standard deviations of their mean values.
This resulted in 49 good radial velocity points over a total time-frame of 730 days. The individual radial velocity data points are given in Table~\ref{tab:vel}.

\begin{figure}
\resizebox{\hsize}{!}{\includegraphics{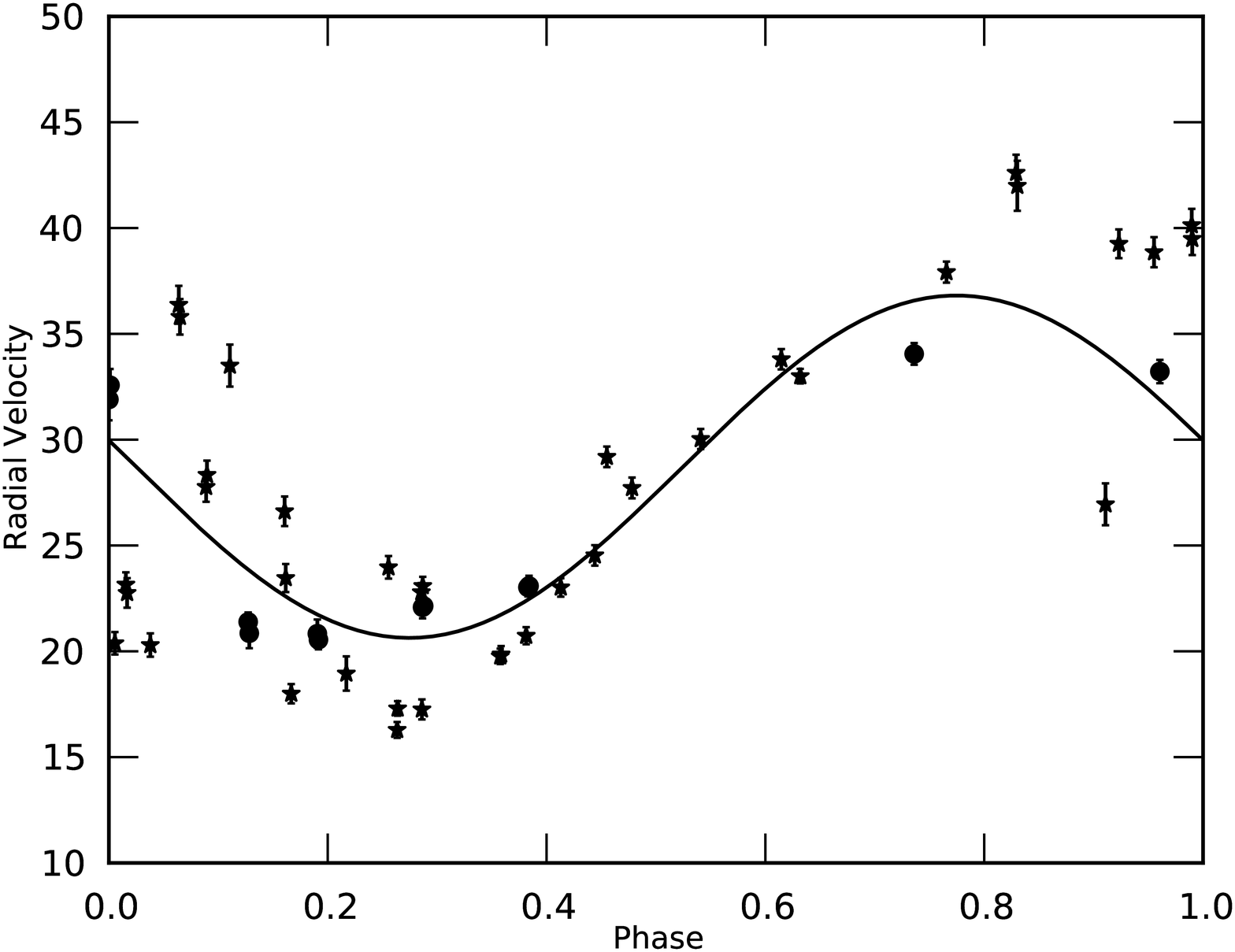}}
\caption{\label{fig:phaseVR} 
The radial velocity data folded on the pulsation period of 31.16 days. 
The solid line shows a sine curve through a single, well-sampled cycle for which the data 
are plotted as filled circles. All other data are plotted as filled stars.}
\end{figure}

\onltab{6}{
\begin{table*}
\caption{\label{tab:vel}
Barycentric radial velocity data of IRAS 11472$-$0800 obtained with the HERMES spectrograph.}
\begin{tabular}{rrrrrr} \hline\hline\rule[0mm]{0mm}{3mm}
HJD - 2400000 & v$_{rad}$  & error  &  HJD - 2400000 &
v$_{rad}$  & error  \\ \hline
day & \kms & \kms &  day & \kms & (\kms) \\ \hline
    55211.7413079  &	 42.61  &	 0.85 	    & 55664.4513216  &	 19.76  & 	 0.36  \\ 	
    55211.7767163  &	 41.99  &	 1.18 	    & 55664.4750508  &	 19.84  &	 0.39  \\	
    55216.7432261  &	 40.13  &	 0.77 	    & 55684.4660032  &	 31.89  &	 0.98  \\	
    55216.7577181  &	 39.48  &	 0.76 	    & 55684.4978335  &	 32.57  &	 0.76  \\	
    55251.6714953  &	 33.50  &	 0.99 	    & 55688.4326792  &	 21.38  &	 0.44  \\	
    55293.5673953  &	 29.18  &	 0.48 	    & 55688.4645092  &	 20.85  &	 0.70  \\	
    55298.5359750  &	 33.80  &	 0.48 	    & 55690.4026638  &	 20.83  &	 0.66  \\	
    55319.4441090  &	 22.78  &	 0.43 	    & 55690.4344930  &	 20.54  &	 0.45  \\	
    55319.4838188  &	 23.08  &	 0.43 	    & 55693.3987808  &	 22.08  &	 0.52  \\	
    55334.3990715  &	 37.91  &	 0.49 	    & 55693.4306107  &	 22.13  &	 0.40  \\	
    55568.7406503  &	 17.25  &	 0.47 	    & 55696.3990050  &	 23.02  &	 0.43  \\	
    55571.7096404  &	 20.73  &	 0.40 	    & 55696.4308346  &	 23.08  &	 0.48  \\	
    55572.6937496  &	 23.01  &	 0.43 	    & 55707.3988811  &	 34.05  &	 0.51  \\	
    55573.6638451  &	 24.53  &	 0.48 	    & 55714.3965047  &	 33.21  &	 0.55  \\	
    55574.7235709  &	 27.71  &	 0.49 	    & 55718.3949237  &	 27.76  &	 0.70  \\	
    55576.6803859  &	 30.03  &	 0.47 	    & 55718.4198083  &	 28.33  &	 0.67  \\	
    55610.6751655  &	 33.00  &	 0.34 	    & 55722.3882184  &	 18.94  &	 0.81  \\	
    55622.6297044  &	 23.15  &	 0.58 	    & 55744.3893212  &	 39.25  &	 0.67  \\	
    55622.6646696  &	 22.75  &	 0.69 	    & 55745.3902829  &	 38.85  &	 0.71  \\	
    55650.5283546  &	 26.94  &	 0.99 	    & 55935.7368893  &	 36.37  &	 0.89  \\	
    55653.4768158  &	 20.37  &	 0.53 	    & 55935.7687243  &	 35.80  &	 0.84  \\	
    55654.4859780  &	 20.29  &	 0.55 	    & 55938.7509075  &	 26.61  &	 0.69  \\	
    55658.4988853  &	 17.99  &	 0.45 	    & 55938.7822517  &	 23.46  &	 0.66  \\	
    55661.5154086  &	 16.28  &	 0.37 	    & 55941.7087357  &	 23.96  &	 0.53  \\	
    55661.5287215  &	 17.29  &	 0.34 	    &		     &		 &	        \\\hline
\end{tabular}
\end{table*}
}

\begin{figure}
\resizebox{\hsize}{!}{\includegraphics{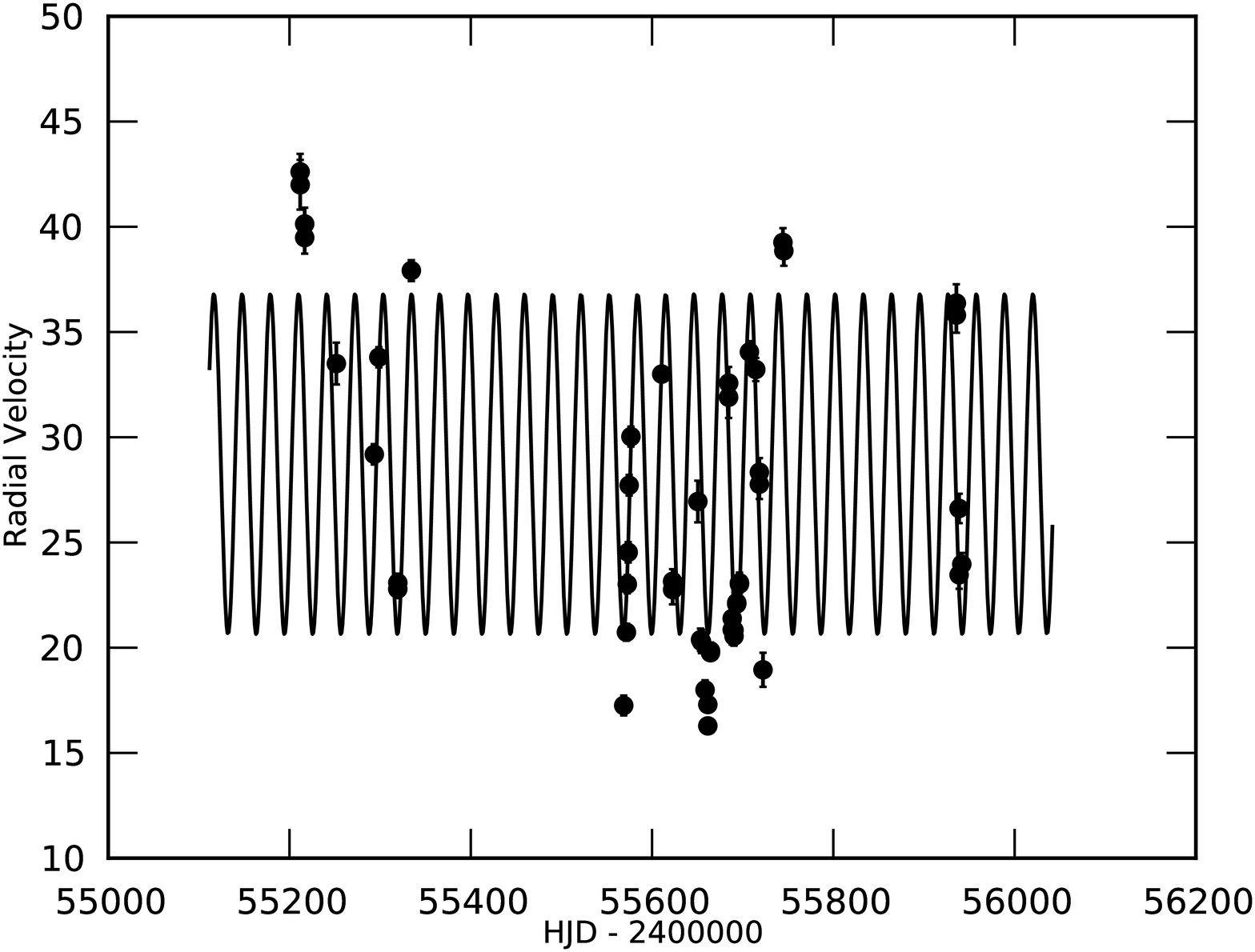}}
\caption{\label{fig:timeVR} 
The radial velocity data of \iras\,. The solid line in the sine curve determined by the fit to the single, well-sampled cycle.}
\end{figure}

The object is clearly variable in radial velocity, and in
Fig.~\ref{fig:phaseVR} we show the data folded on the pulsation
period of 31.16 days. The sine curve is based on the fit of a single, isolated,
well-sampled cycle, the data of which are indicated with a different
symbol on the plot. The signature of the pulsation is clearly visible in the
radial velocity data, but the cycle-to-cycle variability is significant.
This is also seen in Fig.~\ref{fig:timeVR}, where the same sine curve fit is compared to
the radial velocity data plotted over time.  The photometric period appears to be a good fit,
but some of the data in the different seasons fall systematically outside the pulsation curve model.

\begin{figure}
\resizebox{\hsize}{!}{\includegraphics{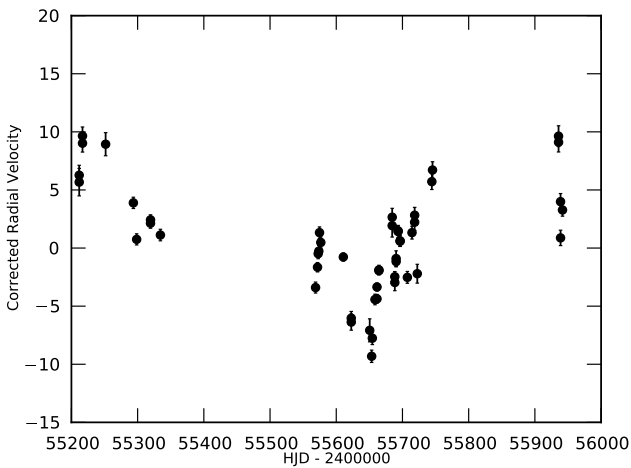}}
\caption{\label{fig:residualVR} 
The residual radial velocity data of \iras\, after subtraction of the sine curve shown in Fig.~\ref{fig:timeVR}.}
\end{figure}

This is seen more clearly when one removes the sine curve and examines the residuals, 
as shown in Fig.~\ref{fig:residualVR}.
They suggest longer time scale systematic effects, perhaps of a cyclical nature.  Although the data
do not cover an entire cycle, we think that they are likely due to orbital motion,
based on the similarity of this star to others that have been found to be binary.

\begin{figure}
\resizebox{\hsize}{!}{\includegraphics{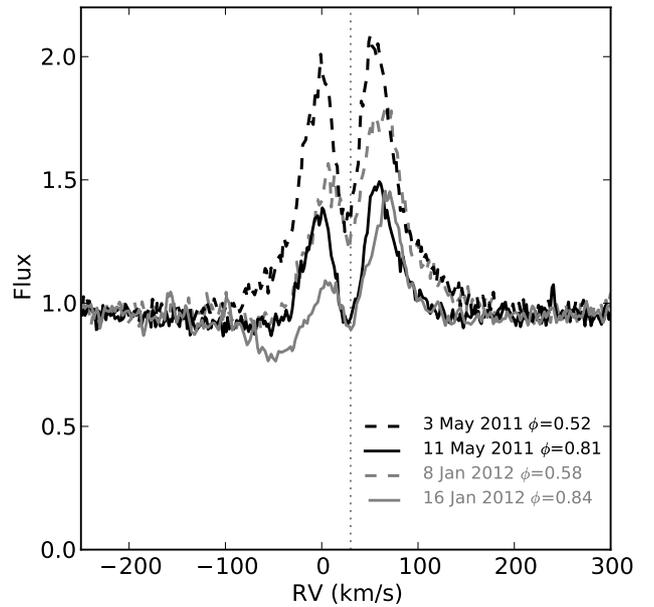}}
\caption{\label{fig:ha} Continuum normalised H$_{\alpha}$ line
  profiles. The indicated phases are computed from the ephemerides
  given in Sect. 4.}
\end{figure}

Continuum-normalized representative profiles of H$\alpha$ are show in Fig.~\ref{fig:ha}.
The dotted line marks the systemic velocity.
The strength of the double-peak
emission correlates with the pulsation phase, but significant
cycle-to-cycle variations are also observed,
which may be related to the orbital motion.

The confirmation and determination of the orbital period of \iras\,
will need a significantly longer time series with appropriate
sampling. The current season has just started and we will continue to
monitor this object.

\section{Discussion}

\begin{figure}
\resizebox{\hsize}{!}{\includegraphics{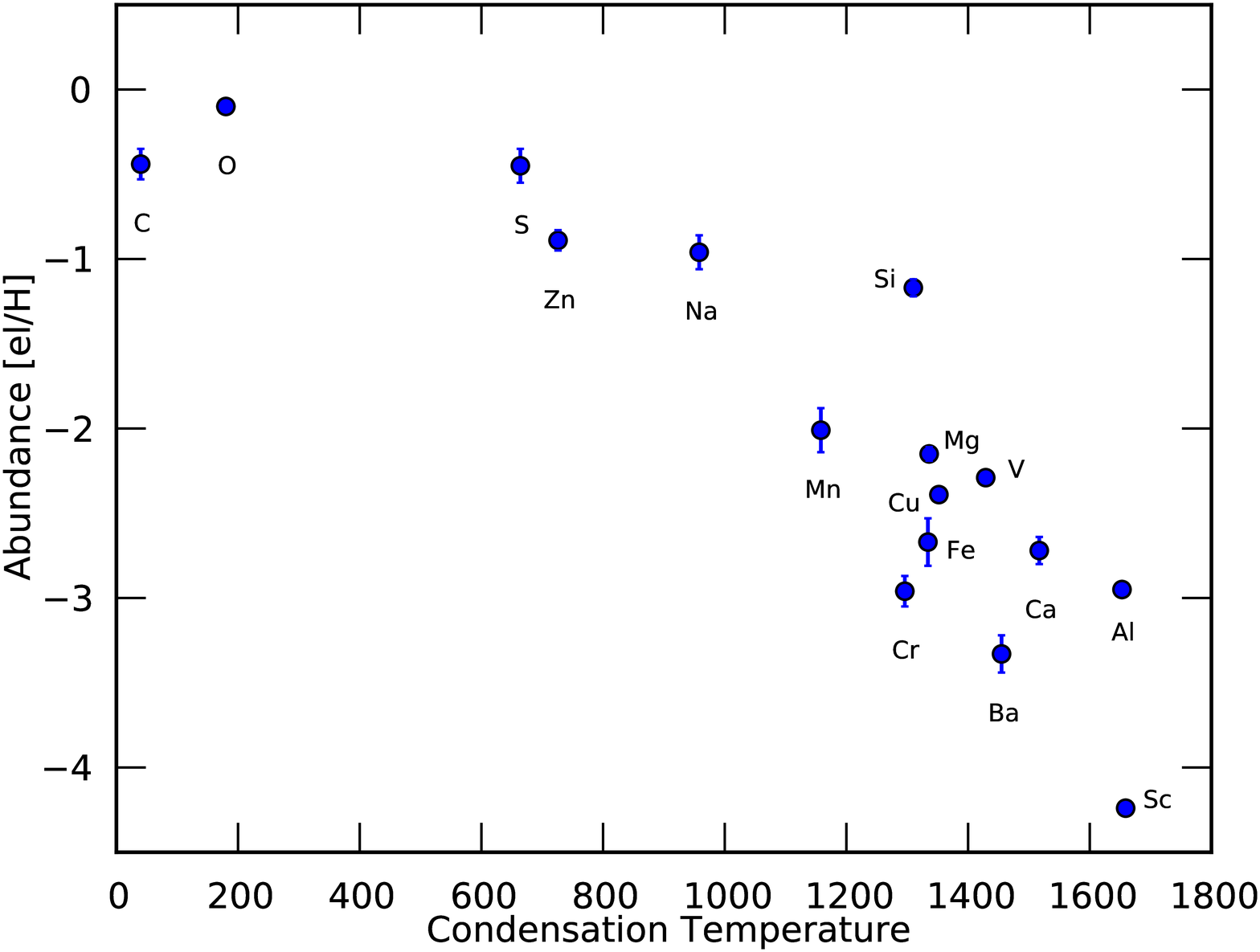}}
\caption{\label{fig:Tcond} 
The photospheric abundances of \iras\, relative to the condensation
temperature of the chemical element. The condensation temperatures are
from \cite{lodders03}}
\end{figure}

With [Fe/H]=$-$2.7, [Cr/H]=$-$3.0, [Sc/H]=$-$4.2, [Ba/H]=$-$3.3
and the strong correlation of the abundances with dust condensation
temperature (Fig.~\ref{fig:Tcond}), \iras\, is one of the most
depleted objects known to date!

Another outstanding property of this object is the SED
(Fig.~\ref{fig:sed}), showing that the thermal emission of the
circumstellar dust dwarfs in luminosity the dereddened photospheric emission.
The infrared excess is warm and the silicate spectrum is very similar to the strongly processed silicates analysed
by \citet[][and references therein]{gielen11}. The warm excess, peaking at about $\sim$ 300K and
coming from dust close to the star, and also the spectral properties of the grains are
clear indications that the circumstellar dust is trapped in a disc.
Given the energetics, our aspect angle towards the disc is likely close
to edge-on and the small optical flux we detect may very well be dominated by scattered
light. Evolved objects with trapped dust in a disc and similar viewing angles are the Red
Rectangle \citep[e.g.][]{cohen75, cohen04}, IRAS20056+1834
\citep[e.g.][]{menzies88, rao02} and IRAS17233-4330
\citep[e.g.][]{deruyter06,gielen08}, which are indeed all sources where
the infrared luminosity dominates over the dereddened
photometry. However, apart from the Red Rectangle, the other objects are much
less chemically depleted if at all.

It is by now well known that the stable disc environment facilitates
strong dust grain processing, and \iras\, is no exception. We lack
a wide spectral coverage of our IR spectroscopy.  To model the
spectrum, we
employ the same technique as \cite{gielen08,gielen11},
and assume that the silicate feature can be decomposed in the optically-thin
regime by adding the spectral contributions of different dust species
with a range of grain sizes. We limited our analysis using three
grain sizes (0.1/2/4 $\mu$m). Our decomposition shows (Fig.~\ref{fig:timmi}) that
the silicate feature is dominated by large amorphous grains with a
significant contribution of crystalline forsterite, which is
responsible for the clear 11.3 $\mu$m signature. The presence of these
highly processed silicate grains is a mainstream characteristic of
disc sources with evolved central stars \citep[e.g.,][and references
therein]{gielen11} and \iras\, proves not to be an exception.

Further evidence for a compact, optically-thick disc seen nearly
edge-on is found in the near-infrared
imaging polarimetric study by \citet{gledhill05}.  The object is found to
be strongly polarized, with a maximum polarization of 10~\% in J and
8~\% in K and integrated polarization of 6~\% in each; since it is out
of the galactic plane ($b$ = +51.6\degr) and stars in the same
direction have little polarization, this is judged to be essentially
entirely intrinsic.  The source is unresolved with a polarisation 
pattern like that arising from a small disc and with a strong
scattering component.

The system has the unusual property of being bluer in general when it
is fainter, a condition not expected if the only mechanism operating
is extinction by dust.  However, also this can be understood if the
extinction is accompanied by scattering into the line of sight which
comes to dominate the colour even more when the object is globally
fainter.  Such a change in colour of becoming bluer when fainter is
seen in some Herbig Ae stars and is suggested indeed to be due to dust
scattering.  \citet{bibo90} derive a model in which extra blue light
is derived from scattering by fine dust in the circumstellar envelope
and the overall dimming of the light is due to orbiting opaque dust
clouds.  In a variation of this model, it is the orbit of the star
itself that changes the overall extinction and dimming of the direct
light rather than orbiting dust clouds.  If the star is orbiting
within a disc, then one can easily imagine a phase dependence in the
brightness as light received passes though differing amounts of the
dust.  However, if this is a uniform disc, then one would expect to
see the system return to its earlier brightness level. If the very
long timescale as observed in the photometric monitoring is indeed
linked to an orbital period, this period is in the order of 14 years
or longer.  This is significantly longer than the orbital periods
found for measured post-AGB stars found binary systems, which range
from 100 to 3000 days \citep{vanwinckel07,vanwinckel09}.

The radial velocity variations are dominated by the pulsations, but we
did discover a longer term trend which we interpret as due to orbital
motion. The interpretation of this longer term trend is not straightforward,
as the measured velocity, is not the radial compoment of the velocity
of the star due to the dominant scattered light. It is therefore
difficult to de-project the amplitude of the radial velicity to the
putative orbital plane. We will
continue to monitor this source to discover, hopefully, the
long-term period as well as the orbital elements of the binary.

The pulsation period and spectral properties put \iras\, in the realm
of the RV\,Tauri stars, despite the fact that we do not find 
evidence in the light curve for the characteristic succession of deep
and shallow minima. Similar RV\,Tauri pulsators do exist, however, and
a particular good example is MACHO14.9582.9 in the LMC
(J053932.79-712154.4) with a very similar period (31.127 d.) and
light curve \citep{alcock98}. Also this object has a clear infrared
excess but the aspect angle is different as can be seen from the SED
\citep{vanaarle11}.  The RV\,Tauri stars form a PL relation of their
own \citep{alcock98, buchler09}, and when we apply the PL relation of
the LMC and use MACHO14.9582.9 as a proxy of \iras\,, we deduce a
luminosity of $\sim$2000 L$_{\odot}$. Adopting the total integral of
the raw SED of \iras\, as a good estimate of the total luminosity, 
we obtain a distance of about 2 kpc for this source.

We conclude that the hitherto poorly studied \iras\, is a strongly
depleted evolved star, which is surrounded by a stable dusty disc. The
object is a regular pulsator with a period of 31.16 d which is in the
regime of the population II Cepheids at the lower luminosity tail of
the RV\,Tauri stars. The light curve does not show the alternation of
deep and shallow minima which is characteristic of RV\,Tauri objects.
The energetics of the SED as well as the colour behaviour of the 
long-term trend in the multi-colour photometry shows that the viewing angle
to the system is close to edge-on, which means that the optical flux is
dominated by scattering. The pulsations are well recovered in our
radial velocity time-series, and we interpret the 
systematic residual velocities as due to orbital motion. 
The orbital parameters are not yet determined. 
With its low luminosity and regular systematic period of 31.16d, we
conclude that \iras\, is a low-luminosity analogue of the dusty
RV\,Tauri stars.

\begin{acknowledgements}
  We thank W. Zima for a help discussion about period analysis and
  using Period04.
  HVW acknowledges financial support from the Research Council of
  K.U.Leuven under grant GOA/2008/04 and from the Scientific Fund of
  Flanders (FWO) under the grants G.0703.08 and G.0470.07. 
   BJH acknowledges ongoing financial support from the National
  Science Foundation (most recently through AST 1009974). 
  Long-term monitoring is only possible thanks to the dedicated efforts of many
  observers. The authors want to acknowledge following people who
  contributed to obtaining the HERMES data: Bram Acke, Fabio Barblan,
  Steven Bloemen, Nick Cox, Peter De Cat, Pieter Degroote, Ben
  Devries, Laurent Eyer, Yves Fremat, Patricia Lampens, Robin
  Lombaert, Pieter Neyskens, P\'eter P\'apics, Kristof Smolders, Andrew
  Tkachenko, Christoffel Waelkens as well as the whole local Mercator team.  
  We also acknowledge the observing contribution of
  many undergraduate research students at Valparaiso University. 
  This research has made use of the SIMBAD database, operated at CDS,
  Strasbourg, France, and NASA's Astrophysics Data System.

\end{acknowledgements}

\end{document}